# A scoping review of using Large Language Models (LLMs) to investigate Electronic Health Records (EHRs)

Lingyao Li*[1], Jiayan Zhou*[2], Zhenxiang Gao[3], Wenyue Hua[4], Lizhou Fan[1], Huizi Yu[5], Loni Hagen[6], Yongfeng Zhang[4], Themistocles L. Assimes[2], Libby Hemphill[1], Siyuan Ma[7]

[1] School of Information, University of Michigan
[2] School of Medicine, Stanford University
[3] School of Medicine, Case Western Reserve University
[4] Department of Computer Science, Rutgers University
[5] School of Public Health, University of Michigan
[6] School of Information, University of South Florida
[7] Department of Biostatistics, Vanderbilt University Medical Center

**Abstract:** Electronic Health Records (EHRs) play an important role in the healthcare system. However, their complexity and vast volume pose significant challenges to data interpretation and analysis. Recent advancements in Artificial Intelligence (AI), particularly the development of Large Language Models (LLMs), open up new opportunities for researchers in this domain. Although prior studies have demonstrated their potential in language understanding and processing in the context of EHRs, a comprehensive scoping review is lacking. This study aims to bridge this research gap by conducting a scoping review based on 329 related papers collected from OpenAlex. We first performed a bibliometric analysis to examine paper trends, model applications, and collaboration networks. Next, we manually reviewed and categorized each paper into one of the seven identified topics: named entity recognition, information extraction, text similarity, text summarization, text classification, dialogue system, and diagnosis and prediction. For each topic, we discussed the unique capabilities of LLMs, such as their ability to understand context, capture semantic relations, and generate human-like text. Finally, we highlighted several implications for researchers from the perspectives of data resources, prompt engineering, fine-tuning, performance measures, and ethical concerns. In conclusion, this study provides valuable insights into the potential of LLMs to transform EHR research and discusses their applications and ethical considerations.

*These two authors contribute equally to this paper.

**Author Emails**: lingyaol@umich.edu, jyzhou@stanford.edu, zxg306@case.edu, wenyue.hua@rutgers.edu, lizhouf@umich.edu, huiziy@umich.edu, lonihagen@usf.edu, yongfeng.zhang@rutgers.edu, tassimes@stanford.edu, libbyh@umich.edu, siyuan.ma@vumc.org.

# 1 Introduction

Electronic Health Records (EHRs) are a critical component of the modern healthcare system, transforming how patient information is stored, accessed, and utilized[1]. These records are digital repositories of patients' medical history that encompass a wide range of patient information[2,3]. Over the past decade, the adoption of EHRs by hospitals has witnessed a dramatic increase, partly driven by the Health Information Technology for Economic and Clinical Health (HITECH) Act of 2009[4,5]. The nationwide adoption rate of EHR systems by hospitals has increased from 6.6% in 2009 to 81.2% in 2019[6]. The wealth of information in EHRs has opened up new avenues for researchers and healthcare professionals to conduct retrospective cohort studies[7,8] and identify a broad range of disease outcomes and predictors, such as disease phenotyping[9], patient selection for clinical trial[10], and rare disease diagnosis[11,12].

Nevertheless, fully leveraging the potential of EHRs has been a significant challenge[13,14]. One primary limitation is the lack of qualitative and accurate information on disease phenotype. The diagnostic billing codes and the International Classification of Diseases (ICD) codes, which are often used to identify patient cohorts as standardized coding methods, may not accurately reflect patients' current disease statuses[15,16]. These issues can arise due to uncertainties in diagnosis, particularly in the early phase of the disease, as well as inconsistent coding practices[17]. Such presence of incompleteness and inconsistency also makes it difficult to compare and aggregate EHR data across different healthcare systems and providers[18]. In addition, the complexity and vast volume of EHR data presents significant challenges in data analysis, especially when dealing with unstructured data such as clinical notes[19]. This nature makes manual annotation and interpretation impractical and time-consuming.

Natural Language Processing (NLP) techniques enable efficient processing of clinical notes and narratives and have been extensively used to process EHR data[20,21]. One popular area aims to enhance clinical documentation by extracting useful information from both structured data (e.g., lab test results and vital signs) and unstructured data (e.g., clinical notes) in EHRs. NLP techniques, such as named entity recognition[22,23] and text summarization[24], have been employed to extract entities relevant to patient health status and summarize treatment outcomes. The other area focuses on downstream applications of EHR data, such as the examination of disease progression[25] and adverse drug reactions[26]. These studies can help enhance the understanding of medical conditions and therapeutic interventions.

Despite their wide applications, conventional NLP techniques suffer several limitations. One limitation lies in their ability to interpret the intricate and subtle nuances inherent in medical language. This challenge arises due to the prevalence of jargon, abbreviations, and context-specific terminologies in medical language[27] within EHRs. For example, jargons often convey hidden meanings accepted and understood depending on the specific field, which could be challenging for the general NLP techniques to interpret[28]. In addition, due to discrepancies in documentation practices and patient populations, NLP models trained on data from one healthcare institution often struggle to generalize effectively to data from other institutions. As a result, training NLP models can require a substantial amount of human annotation[29]. Further, conventional NLP techniques often lack general-purpose capabilities[30] as they are trained for specific tasks. This can make it challenging to assist with patients' diverse range of questions and concerns, such as interpreting medical jargon and understanding symptoms. These inherent limitations of NLP techniques underscore the need for more robust and generalizable tools to process information across diverse healthcare settings.

Large Language Models (LLMs) have recently emerged as novel technologies for language processing. These models leverage deep neural networks with billions of parameters, trained on gigantic amounts of unlabeled text data through self-supervised learning[31,32]. Compared to conventional NLP techniques, LLMs exhibit generative capabilities to comprehend contextual language and generate human-like text across a broad set of NLP tasks[32,33]. These enable LLMs to interpret the meaning of text even for ambiguous or complex language structures that commonly appear in HER data. Typical LLMs include Bidirectional Encoder Representations from Transformers (BERT) developed by Google[34], Generative Pre-trained Transformer (GPT) families developed by OpenAI[35], and Large Language Model Meta AI (LLaMA) by Meta[36]. Since their release, researchers have demonstrated the potential of using LLMs for EHR data to enhance clinical documentation and support decision-making[37–40].

With the latest release of LLMs such as GPT4 and Claude3, the application of LLMs in biomedical and health informatics has become a highly sought-after research area[41,42]. While previous literature has emphasized the potential of LLMs in the domain of biomedical informatics[43–45], a scoping review specifically focusing on EHRs remains absent. A scoping review can help to show the current landscape of LLM applications in EHRs, as well as provide insights for researchers into how to use different LLMs and techniques to advance EHR data analysis and further improve clinical practices. Therefore, this paper aims to address this gap by presenting a scoping review of current studies encompassing various research topics using LLMs to process EHR data. The significance of this paper is multifaceted, which includes,

1. Present an up-to-date bibliometric analysis of current studies and identify trends and patterns that deepen understanding of LLMs' applications.
2. Categorize current research into seven general topics and discuss relevant potential and challenges within each topic.
3. Provide guidance for researchers and practitioners seeking to leverage LLMs for using EHR data and identify research opportunities.

## 2 Materials

**Figure 1** presents a visual illustration of the research roadmap. In this review study, we first introduce EHR data and data types (**Section 2.1.1**), followed by a summary of challenges associated with EHR data (**Section 2.1.2**). Next, we provide a brief introduction to LLMs, including the model structure (**Section 2.2.1**), fine-tuning mechanism (**Section 2.2.2**), and in-context learning (**Section 2.2.3**). Based on the knowledge of both EHRs and LLMs, we compile a list of search terms to collect relevant studies from OpenAlex (**Section 2.3**). Upon the collection, we perform a bibliometric analysis, as presented in **Section 3.1**. Subsequently, we categorize studies into seven identified topics and specifically discuss each of them from **Sections 3.2.1** to **3.2.7**.

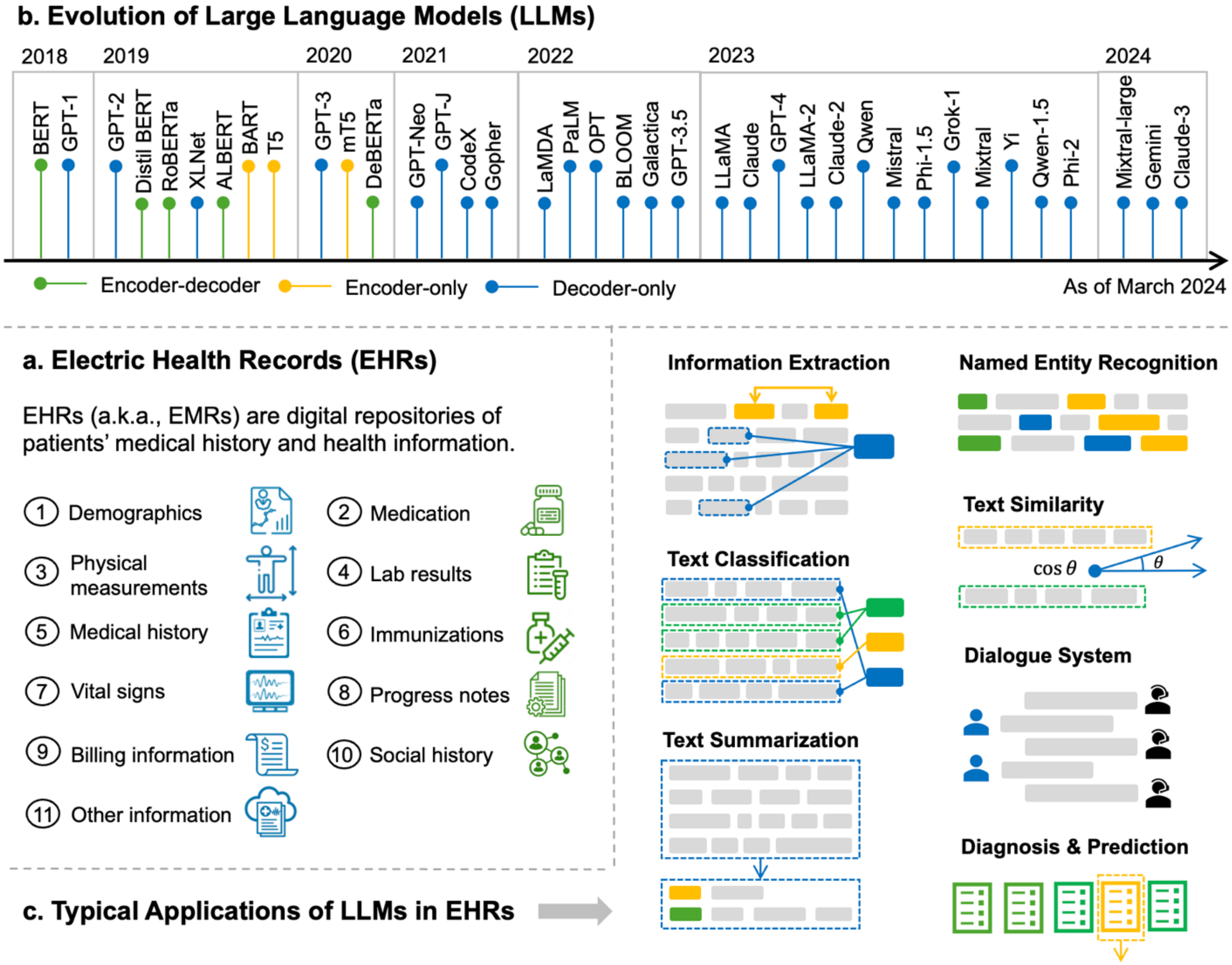

**Figure 1**. An illustration of the research roadmap: (a) EHR data, (b) Evolution of LLMs, and (c) Typical applications of LLMs in EHRs.

## 2.1 Introduction to EHRs

### 2.1.1 EHR data and types

EHRs are digital repositories of patients' medical history and health information[46]. The term EHRs is also known as electronic medical records (EMRs). The distinction is that EHRs are maintained by multiple providers, while EMRs are maintained by a single provider. However, both terms refer to electronic patient records and are often used interchangeably. These records encompass a wide range of data crucial to patient care, including but not limited to, demographics, medications, physical measurements, lab results, medical history, immunizations, vital signs, progress notes, billing information, and social history[47] (as illustrated in **Figure 1(a)**). EHR systems can help automate patient information and generate complete records of patient encounters, thus facilitating or improving clinical practice including evidence-based decision-making, quality management, and outcomes reporting[48]. They have also shown significant potential to enhance patient care, improve the clinician experience, and provide valuable information for biomedical research[49].

EHR data can be categorized into three major types: (1) structured, (2) semi-structured, and (3) unstructured data, each serving a distinct role in patient care and healthcare management[50].

- **Structured data**, the most easily quantifiable and searchable, is stored in predefined formats within databases[51]. It includes essential patient information such as birth date, nationality, prescribed medications, allergies, and key vital signs like height, weight, blood pressure, and blood type.
- **Semi-structured data** offers a balance between rigid structure and flexible narration, typically organized in the flow chart format, similar to RDF (resource description files)[51]. This format captures elements like disease names, test values, and timestamps.
- **Unstructured data**, the most voluminous and narrative-rich, comprises clinical notes, surgical records, discharge summaries, radiology, and pathology reports[52]. This type of data, although challenging to standardize and analyze, provides in-depth insights into patient histories, diagnostic reasoning, and treatment outcomes, making it invaluable for patient care and medical research.

### 2.1.2 Typical EHR data sources

**MIMIC(-II/-III/-IV)** – Medical Information Mart for Intensive Care (MIMIC)[1] is a freely available de-identified medical-related data that was developed and released by the Massachusetts Institute of Technology (MIT) Laboratory for Computational Physiology in collaboration with Beth Israel Deaconess Medical Center[53–55]. It contains a wide range of information from the clinics, including demographics, vital signs, laboratory results, medications, clinical notes, monitoring data, procedures and interventions, and diagnostic codes. Multiple MIMIC datasets are released with periodic updates to include more recent collections, which are designed to support a wide range of research in medical care. The MIMIC-III dataset was utilized most frequently in research employing LLMs to improve current healthcare which contains data from 2001 to 2012. Moreover, MIMIC-IV is the latest version which includes data collected from 2008 to 2019. The specific subsets of data are also available within the MIMIC-IV database, such as hospital-level data (HOSP), ICU-level data (ICU), emergency department data (ED), patient identifiers to link to x-ray data (CXR), ECG data (ECG), and deidentified free-text clinical notes (Note).

**CCKS** – China Conference on Knowledge Graph and Semantic Computing (CCKS)[2] is a conference that is hosted by the Language and Knowledge Computing Professional Committee of the Chinese Information Processing Society of China[56–63]. The conference committee proposed building and sharing the EHR datasets for participants to perform multiple tasks, such as knowledge reasoning, knowledge acquisition, graph computing, and question answering. The data provided by CCKS are all written in Chinese and include Q&A-style data, not limited to EHR data.

**i2b2/n2c2** – National NLP Clinical Challenges (n2c2)[3] is the publicly available dataset that originated from the i2b2 Center, which was an NIH-funded based at Partners HealthCare System in Boston from 2004 to 2014[64]. i2b2 advocated for the potential of clinical records to yield insights for healthcare improvement and provided deidentified notes for NLP Shared Task challenges and workshops. The i2b2 NLP data sets are now released and hosted on the DBMI Data Portal under n2c2. The datasets consist of de-identified clinical notes and texts, with varying topics each year, such as deidentification and smoking in 2006, and deidentification and heart disease in 2014.

[1]MIMIC; https://mimic.mit.edu
[2]CCKS; https://link.springer.com/conference/ccks
[3]n2c2; https://n2c2.dbmi.hms.harvard.edu/

### 2.1.3 Challenges

Drawing insights from prior studies[65–68], there are several major challenges associated with EHRs. First, EHRs often use different **coding systems** such as ICD (International Classification of Diseases), CPT (Current Procedural Terminology), and LOINC (Logical Observation Identifiers Names and Codes) to classify diseases, treatments, and other health-related information. The challenge lies in mapping between these coding systems[69], as inaccurate mappings could reduce the interoperability between disparate healthcare information systems.

The second challenge is associated with the **heterogeneity** in data standards, formats, and terminologies employed by each EHR system[70]. This could hinder accurate medical entity recognition, particularly when dealing with EHRs from diverse vendors. Such intricacies are further exacerbated by variations in medical terminology across different healthcare settings. For instance, a specific medical term may be referred to using different terms or phrases, depending on the location, the medical specialty, or the individual practitioner's preference. Moreover, the routine use of synonyms and abbreviations in EHRs can bring additional complexity to the heterogeneity challenge, as they can be highly variable and context-dependent.

Next, a significant portion of EHR data is composed of **unstructured text**, including clinical notes, lab reports, and imaging data descriptions. This presents a significant challenge when it comes to summarizing or extracting information[71]. The complexity of text summarization from unstructured data arises from the necessity to maintain the clinical accuracy and integrity of the original EHR data while presenting it in a concise and readily accessible format. In addition, the large volume of unstructured text can be overwhelming for information retrieval.

Last, **privacy maintenance** can pose another challenge given that EHRs often contain extremely sensitive information, such as patients' social demographics and medical history. Based on regulatory laws like the Health Insurance and Portability and Accountability Act (HIPAA), maintaining privacy within analytic workflows is imperative[72,73]. As a result, before performing any downstream tasks or sharing data with others, additional privacy-preserving measures must be taken. However, removing private information from a large corpus of EHRs can be costly and challenging to automate, requiring the expertise of annotators with specialized knowledge in the field[65].

## 2.2 Introduction to LLMs

### 2.2.1 Model structure

LLMs are typically categorized into (1) encoder-decoder, (2) encoder-only, and (3) decoder-only. The evolution of these LLMs over time is illustrated in **Figure 1(b)** and **Appendix A.1**. All three types are based on the Transformers structure[74]. The Transformer structure consists of two main parts: the encoder and the decoder, as illustrated in **Figure 2**. The encoder takes an input sequence and generates embeddings for each input token. It is composed of a stack of identical layers, each of which has two sub-layers: a multi-head self-attention mechanism (orange box on the left side in **Figure 2**) and a position-wise feed-forward network (blue box on the left side in **Figure 2**). Each sub-layer is followed by a residual connection and layer normalization (yellow box on the left side in **Figure 2**). The decoder also consists of a stack of identical layers, consisting of a self-attention layer that allows the decoder to attend to its output history ("musked multi-head attention" in **Figure 2**), an encoder-decoder attention layer that allows the decoder to attend to the output of the encoder, and a feed-forward neural network (blue box on the right side in **Figure 2**). In the following sections, we briefly introduce each type and representative LLMs[31,75,76].

- **Encoder-Decoder** models directly use the original structure from the Transformers with a left-to-right decoder to autoregressively generate the output conditioned on a separate encoder for the input text. Representative models include T5[77] and BART[78].
- **Encoder-only** models use the encoder part of the Transformers. Representative encoder-only models include BERT[34] and its derivatives such as RoBERTa[79] which utilizes a dynamic masking pattern, ALBERT[80] which is smaller and faster to train, and SpanBERT[81] which masks spans instead of tokens.
- **Decoder-only** models incorporate the decoder part of the Transformers, which autoregressively generates tokens from left to right by applying a unidirectional attention mask on input tokens. This mechanism can ensure each token focuses on its preceding tokens and itself. The decoder-only structure has gained significant popularity and is currently the most widely adopted model for recently released LLMs, such as the GPT series[82–85] and the LLaMA series[36,86].

Other than the three typical types of LLMs, the Mixture-of-Expert (MoE)[87] structure has recently gained much popularity. The MoE structure consists of multiple networks which are called experts. Each token in the input sequence is initially processed by a gating network, which determines which expert or experts should be activated for that particular input. The gating network produces a set of weights that indicate the relevance of each expert for the input. Based on these weights, only the most relevant experts are activated, and their outputs are combined to produce the final output of the MoE model. This selective activation of experts allows MoE models to be much more efficient than dense models, where all neurons are activated for every input. Several popular LLMs have adopted the MoE structure, such as Mixtral[88] and Grok-1.

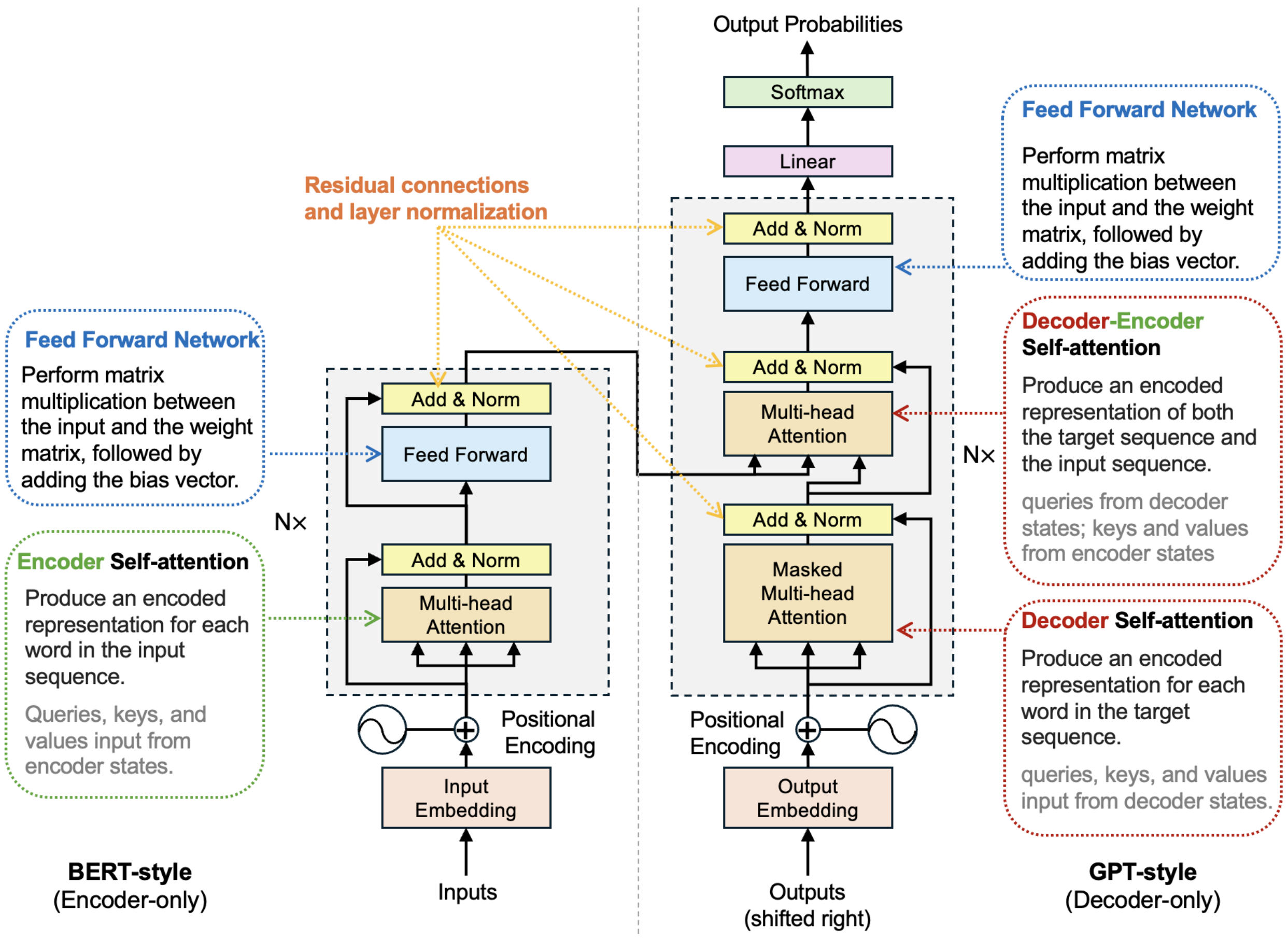

**Figure 2**. The encoder-decoder structure in LLMs (adjusted from Vaswani *et al.*[74]).

### 2.2.2 Fine-tuning

LLMs possess robust capabilities in representing general domain knowledge, yet they often require customization for specific downstream tasks or domains, usually achieved through fine-tuning. In conventional full-parameter fine-tuning, all model weights are modified, presenting inefficiencies, especially for LLMs. Consequently, several parameter-efficient fine-tuning techniques have been developed to improve efficiency, as briefly introduced below.

**Adapter Tunning** – Adapter tuning introduces small and trainable modules called adapters into the layers of a pre-trained model, allowing for fine-tuning to be performed exclusively on these added modules. This approach can significantly reduce the computational cost and preserve the general knowledge of the original pre-trained model. The idea of adopting an adapter for fine-tuning LLM was first proposed by Houlsby *et al.*[89]. In its design, adapter modules (each module consists of two feed-forward layers) are added between each feed-forward layer and the Layernorm layer. During fine-tuning, the pre-trained model remains frozen, and the adapter modules learn the knowledge specific to the downstream task. The addition of adapter modules can mitigate the efficiency issue of fine-tuning the entire model and prevent catastrophic forgetting. Rücklé *et al.*[90] observed that while the adapter presents greater efficiency in training, it slows down at inference time, compared with full-model fine-tuning. Motivated by this, they proposed AdapterDrop, a technique that dynamically removes adapters from lower Transformer layers. This can enable the shared representations at adapters added at low

Transformer layers to improve inference efficiency, especially when conducting multiple task inferences simultaneously.

**Prompt Tuning** – Another parameter-efficient fine-tuning method[91] involves reformulating downstream tasks into a conditional generation task by designing prompts, followed by fine-tuning the model under the condition of these prompts. We focus exclusively on approaches that treat prompts as trainable parameters. Within the trainable prompt framework, prompt vectors are not constrained by the language model's vocabulary embeddings. Instead, they are optimized as continuous variables and concatenated with word type embeddings before being input into the language model.

Numerous models implement this concept in diverse ways. For example, Prefix-tuning, as described by Lester *et al.*[91], introduces a continuous and task-specific vector sequence comprising virtual tokens that are appended to the beginning of the model input. This method keeps the parameters of a pre-trained language model fixed while optimizing only the prefix for a specific task. To facilitate training, prefix-tuning maps the prefix parameters using an MLP layer instead of directly optimizing the prefix parameters. Additionally, it incorporates trainable parameters into all transformer layers. Similarly, P-tuning, presented by Liu *et al.*[92], transforms prompts into learnable embedding layers and adds trainable parameters solely to the input layer. However, P-tuning employs an optional insertion position for virtual tokens, implying that these tokens are not necessarily fixed at the prefix of the input.

**LoRA –** It is worth noting that both adapter-tuning and prompt-tuning may generally result in a decreased throughput of LLMs. This reduction can be attributed to either the introduction of additional parameters or the processing of longer input sequences. To address this challenge, Low-Rank Adaptation (LoRA)[93] employs the concept of reparameterization. Instead of incorporating new modules into the model or modifying the input, LoRA generates a low-dimensional representation to approximate a specific module within the original LLM. This idea comes from the simple hypothesis that trained models are low "intrinsic ranked" (i.e., model parameters contain redundancies). During the tuning phase, only the low-rank representation undergoes adjustment. Upon training completion, these modules are integrated into the original module that the representation aims to approximate. Consequently, the original model structure remains during the inference, thereby preventing any slowdown in inference speed.

**QLoRA**[94] represents an extension of the LoRA approach, operating by quantizing the precision of weight parameters in the pre-trained LLM to 4-bit precision. Generally, trained model parameters are stored in a 32-bit format; however, QLoRA compresses them to a 4-bit format. This compression substantially decreases the memory footprint of the LLM, enabling fine-tuning on a single GPU.

### 2.2.3 In-context learning (ICL)

ICL is a new paradigm for using LLMs[95]. It employs textual inputs to prompt a pre-trained LLM. An LLM then completes tasks by generating responses based on the given instruction and/or a few demonstrations of the task. As the scale of LLMs significantly grows, the expensive cost of fine-tuning LLM makes ICL become a typical approach to utilize LLMs. In the following, we briefly introduce several typical prompt strategies, including zero-shot, few-shot, chain-of-thought prompting, self-consistency prompting, and least-to-most prompting.

**Zero-shot & Few-shot Prompting** – Zero-shot prompting refers to the practice of prompting a pre-trained LLM without incorporating any examples[82]. This approach relies on the LLM's robust capabilities to understand the instruction and complete the task. Few-shot prompting, on the other hand, provides examples within the prompt for the LLM to follow[82]. By offering several demonstrative examples, few-shot prompting can significantly enhance the model's performance compared to zero-shot prompting. Multiple papers have discussed and illustrated the selection of examples to improve few-shot learning performance[33,96–98].

**Chain-of-Thought (CoT) Prompting** – CoT is often designed to enhance LLMs' performance on complex reasoning tasks. CoT provides intermediate reasoning steps to guide the model's responses, which can be achieved through a series of manual demonstrations. Each demonstration consists of a question and a reasoning chain that guides an answer. Prior studies have also explored factors contributing to the effectiveness of CoT and investigated the reasons to explain its mechanism for performance improvement[99,100].

Building upon the concept of CoT prompting, researchers have proposed additional ICL strategies to improve the confidence of reasoning outputs. One such approach is complexity-based prompting[101], which involves conducting a voting process among high-complexity rationales to obtain the final answer. Experimental results demonstrate that prompts with higher reasoning complexity ( i.e., chains with more reasoning steps) can achieve substantially better performance on multi-step reasoning tasks. Another typical one is the Tree-of-Thought (ToT) prompting.[102] This approach integrates tree structures and tree search algorithms into the reasoning construction process, which enables models to efficiently explore and backtrack during the reasoning process to search globally optimal solutions. Moreover, Graph-of-Thought (GoT)[103] introduces more complex topological relationships into prompting. Besides the backtracking operations of ToT, GoT incorporates aggregation and refinement operations to elicit improved reasoning in complex tasks.

**Self-consistency Prompting** – Self-consistency prompting aims to ensure the consistency of each round of responses from LLMs, motivated by the intuition that if a model is asked to answer a series of related questions, the answers should not contradict each other. Wang *et al.*[104] are the first to incorporate self-consistency into prompting LLMs. Their study samples a diverse set of reasoning paths instead of only taking the greedy one and then selects the most consistent answer by marginalizing out the sampled reasoning paths. Using their approach, a complex reasoning problem can admit multiple ways of thinking that lead to its unique correct answer.

**Least-to-most Prompting** – This prompt method is an innovative method proposed by Zhou *et al.*[105], which decomposes intricate problems into a succession of more rudimentary subproblems that can be sequentially addressed. The resolution of each subproblem can be expedited by leveraging solutions derived from antecedent subproblems. This process allows the model to build upon its previous knowledge and gradually progress towards solving the overall complex problem.

## 2.3 Data preparation

We selected OpenAlex[106] to collect related studies for two primary reasons. First, OpenAlex is an open-source repository of scholarly metadata, which enables us to gather both recently archived preprints and published articles from journals and conferences. This is especially helpful for our scoping review, considering the prevalence of preprinted studies in the domain of LLMs. Second, OpenAlex offers its data freely and openly, ensuring our analysis can be replicated within the community without licensing barriers.

Within OpenAlex, we collected relevant papers using two categories of keywords, namely LLMs and EHRs, as outlined in **Table 1**. For LLM-related keywords, we used two general terms, "large language model" and "LLM," along with popular open-source models (e.g., BERT, T5, LLaMA) and closed-source models (e.g., GPT, Claude). It was worth noting that, although some LLMs were listed in **Figure 1(b)**, they were not included in our search queries. Some models like Yi or Phi did not result in relevant papers and therefore could bring significant noise for paper screening, while other models like grok-1 or Galactic did not return any search results. For EHR-related keywords, we selected two search terms, "Electronic Medical Record" and "Electronic Health Record." We applied these search queries to the title and abstract rather than the full text to increase the likelihood that the returned results were relevant. In addition, we limited our search date range from 2018 to 2024 given that the first widely considered LLM, BERT, was introduced in 2018[34]. These search conditions resulted in a collection of 650 articles.

Subsequently, we conducted a manual review of each paper to ensure its relevance. We used the following process to review candidate papers: (1) the study utilizes EHR data resources, (2) the study employs LLMs to process EHR data, (3) the study is a full research article (i.e., is not an editorial or commentary article), (4) the study is written in English, (5) the study is not a duplicate within the collected dataset, as some authors may choose to preprint the paper, which is later accepted by a journal or conference. After applying these criteria, 329 articles[4] remained and were then used for result analysis.

**Table 1**. Data search and filtering conditions.

| Action | Search and filtering conditions |
|---|---|
| Data search | • Search terms: (Large language model / LLM / BERT / RoBERTa / T5 / XLNet / Mistral / Mixtral / Falcon / Qwen / BLOOM / Vicuna / LLaMA / GPT / Claude / Bard / Google PaLM / Gemini) & (Electronic Medical Record / Electronic Health Record)<br>• Search date: 01/01/2018 ~ 03/31/2024<br>• Paper type: articles |
| Data filtering | • Exclude studies that do not use EHR data resource<br>• Exclude studies that do not use LLMs<br>• Exclude studies that are not full research articles<br>• Exclude studies that are not written in English<br>• Exclude studies that are duplicated |

The next steps were to identify NLP tasks the studies conducted using LLMs and to categorize collected papers into those tasks. To do this, we first conducted a review of current review papers in the domain of biomedical and health informatics, as listed in **Table 2**. Prior research has conducted bibliometric or scoping reviews focusing on either LLMs in biomedical and health informatics or NLP in EHRs, but few have examined the potential of LLMs in EHRs. Although there was a lack of consistency in the categorization of NLP tasks across the literature, we observed that these review studies mentioned several overlapping NLP tasks, including text classification, information extraction, named entity recognition, text summarization, dialogue systems, and text generation. These overlapping tasks reflected the common and critical NLP tasks in EHRs.

[4]The detailed paper information can be accessed via https://github.com/casmlab/llm-ehr.

**Table 2.** Representative review studies using NLP in the domain of biomedical informatics.

| Author & Year | Scope | NLP tasks |
|---|---|---|
| Tian *et al.* (2024)[41] | LLM + Bio-Info | Information retrieval, question answering, medical text summarization, information extraction, medical education |
| Yu *et al.* (2024)[44] | LLM + Bio-Info | Model evaluation, information extraction, dialogue system, multilinguality, text generation, education, ethics, multimodality, inference, summarization, sentiment analysis, named entity recognition. |
| Hossain *et al.* (2023)[20] | NLP + EHR | Medical note classification, clinical entity recognition, text summarization, deep learning, transfer learning architecture, information extraction, language translation, and other applications |
| Wang *et al.* (2023)[107] | LLM + Bio-Info | Information extraction (named entity recognition, relation extraction, event extraction), text classification, sentence similarity, question answering, dialogue systems, text summarization, language inference, prediction. |
| Li *et al.* (2022)[65] | NLP + EHR | Classification (text classification, segmentation, word sense disambiguation, medical coding, outcome prediction), word embeddings (concept, patient, visiting), information extraction (named entity recognition, entity linking, relation and event extraction), generation (EHR generation, summarization, language translation), other topics (question answering, phenotyping, knowledge graphs, medical dialogue, multilinguality, interpretability) |
| Houssein *et al.* (2021)[108] | NLP + Bio-Info | Word sense disambiguation, named entity recognition, adverse drug events, information extraction, relation extraction |
| Wu *et al.* (2020)[109] | NLP + Bio-Info | Text classification, named entity recognition, relation extraction, and others (information retrieval, language generation, abbreviation disambiguation, machine translation, spelling correction). |
| Zhao *et al.* (2020)[110] | LLM + Bio-Info | Medical named entity recognition, text classification, relation extraction, pathway extraction |
| Shickel *et al.* (2017)[5] | NLP + EHR | Information extraction, representation learning, outcome prediction, phenotyping, and deidentification |

We annotated each paper to indicate the NLP tasks it included. This involved a pair-coding process, wherein two authors collaboratively engaged in the reading and annotation of each paper. We employed a pair-coding approach due to its efficacy in facilitating clarification of any ambiguities and expediting the determination of NLP tasks. Upon the conclusion of the initial round of pair coding, a preliminary list of NLP tasks was proposed to the entire team. Later, we held a group discussion to finalize the list of NLP tasks. As a result, we summarized seven major tasks (as illustrated in **Figure 1(c)**), including (1) named entity recognition, (2) information extraction, (3) text summarization, (4) text similarity, (5) text classification, (6) dialogue system, (7) diagnosis and prediction, and (8) others (e.g., translation). Then, a second round of pair coding was conducted to confirm the NLP task for each paper. This iterative process ensured the accuracy and consistency of our paper annotation.

We further conducted a bibliometric analysis to quantify the impact of scholarly works within this field. To prepare for the bibliometric analysis, we collected the following information in addition to the downloaded details (e.g., authors, institutions, journals, citations) from OpenAlex of each paper, including EHR data resources, LLMs, and performance measures. Then, we analyzed the paper trends, data resources, and types of LLMs used in the field. This allowed us to identify the commonly used LLMs and data sources, as well as the overall trends. Next, we examined the collaborative network to gain insight into the overall social status of the research field. This analysis facilitated our understanding of scholarly communication and knowledge diffusion that have contributed to the advancement of the field.

# 3 Results

The results section consists of two subsections. In **Section 3.1,** we conducted a bibliometric analysis, as illustrated in **Figure 3** and **Figure 4**, to highlight the trends in published papers, EHR data resources, the distribution of LLMs, and the institutional collaboration network. In **Section 3.2**, we specifically discussed the applications of LLMs in EHRs regarding each of the seven identified NLP tasks. **Table 3** provides a list of representative studies based on our scoping review.

## 3.1 Bibliometric analysis

**Paper trends** – As **Figure 3(a)** shows, the application of LLMs for EHR research began to emerge in the second quarter of 2019, with a consistent increase corresponding to the growing general interest in LLMs. It is worth noting that after OpenAI released ChatGPT on November 31, 2023, the number of manuscripts showed a dramatic increase in every quarter. In particular, the first quarter of 2024 showed over 50 publications in this area. This upward trend suggests the rapidly expanding interest in using LLMs for EHR research.

**Data sources** – **Figure 3(b)** highlights several EHR data resources that have been widely used in the current studies, including CCKS, n2c2/i2b2, MIMIC, academic medical centers, and hospitals and clinics. Among these, many studies have used EHR data from academic medical centers or private and public hospitals and clinics. On the other hand, several publicly available datasets, such as MIMIC and CCKS, have also been extensively used in EHR studies. In particular, MIMIC-III has been used by 46 studies in our dataset.

**Model trends** – Encoder-only LLMs, represented by the blue dotted line in **Figure 3(c)**, are the most prevalent models in EHR studies. In particular, during 2020 and 2023, encoder-only LLMs have been utilized in the majority of studies, while the other two LLM types have seen minimal employment. Among the encoder-only LLMs, BERT has garnered the most frequent usage. Nearly all related studies have used BERT or BERT-based trained models (**Figure 3(d)**), such as RoBERTa and ClinicalBERT. Another interesting observation is the significant increase in the utilization of decoder-only models following the release of ChatGPT by OpenAI. In the first quarter of 2024, a markedly higher number of studies employed decoder-only models compared to encoder-only models, as the orange line is located higher than the blue dotted line in **Figure 3(c)**. Among those decoder-only models, GPT4 and GPT3.5 are the two most frequently used models, followed by GPT2 and LLaMa2 (**Figure 3(f)**). Encoder-decoder models, such as T5, have seen limited usage in EHR studies but exhibited a modest increase in use in 2024.

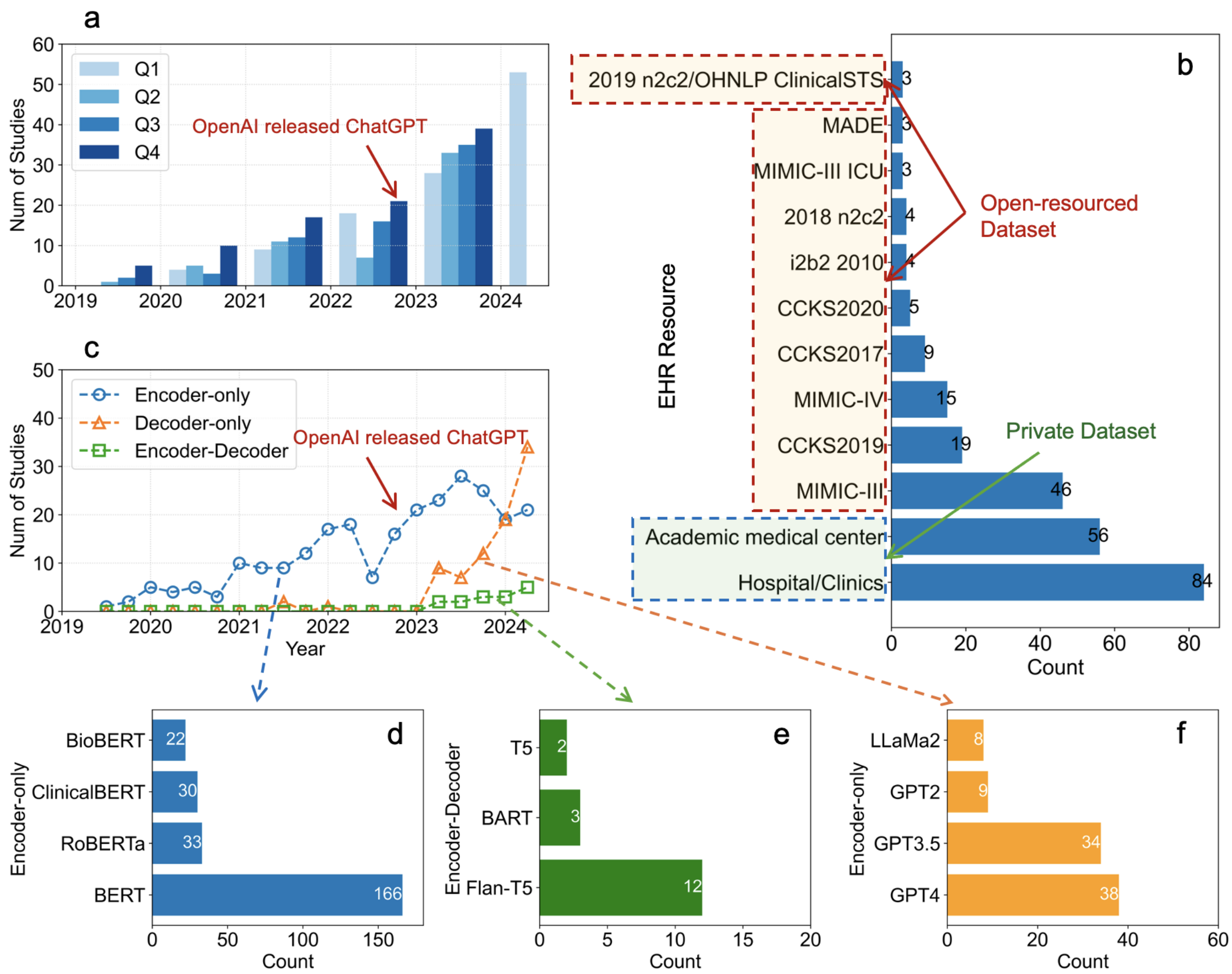

**Figure 3**. Bibliometric analysis based on 329 relevant papers: (a) Paper trend over time from 2019 to 2024 Q1, (b) Count of EHR data resources, (c) LLM adoption over time from 2019 to 2024 Q1, (d) Top four encoder-only LLMs, (e) Top three encoder-decoder LLMs, and (f) Top four decoder-only LLMs.

**Institutional collaboration network** – We used a network analysis tool called VOSviewer[111] to generate the institutional collaboration network based on the authors' affiliated institutions, as presented in **Figure 4**. This visualization can help facilitate a useful perspective on scholarly communication and the dissemination of knowledge in the context of LLMs and EHRs. There are several noteworthy observations. First, several prestigious universities are positioned at the center of the network, including Harvard University (**Figure 4(a)**), Stanford University (**Figure 4(b)**), and Peking University ((**Figure 4(c)**). These institutions are well known for their developed programs in the areas of medicine, public health, and AI. Second, it is evident that many collaborations exist between universities and medical centers or hospitals, such as the collaboration between Harvard University and Boston Children's Hospital and Brigham and Women's Hospital (**Figure 4(a)**), as well as the partnership between Peking University and Peking University People's Hospital ((**Figure 4(c)**). Last, the majority of institutions represented in the network are from the United States and China. This observation suggests that researchers from these two countries are at the forefront of research on LLMs in EHRs.

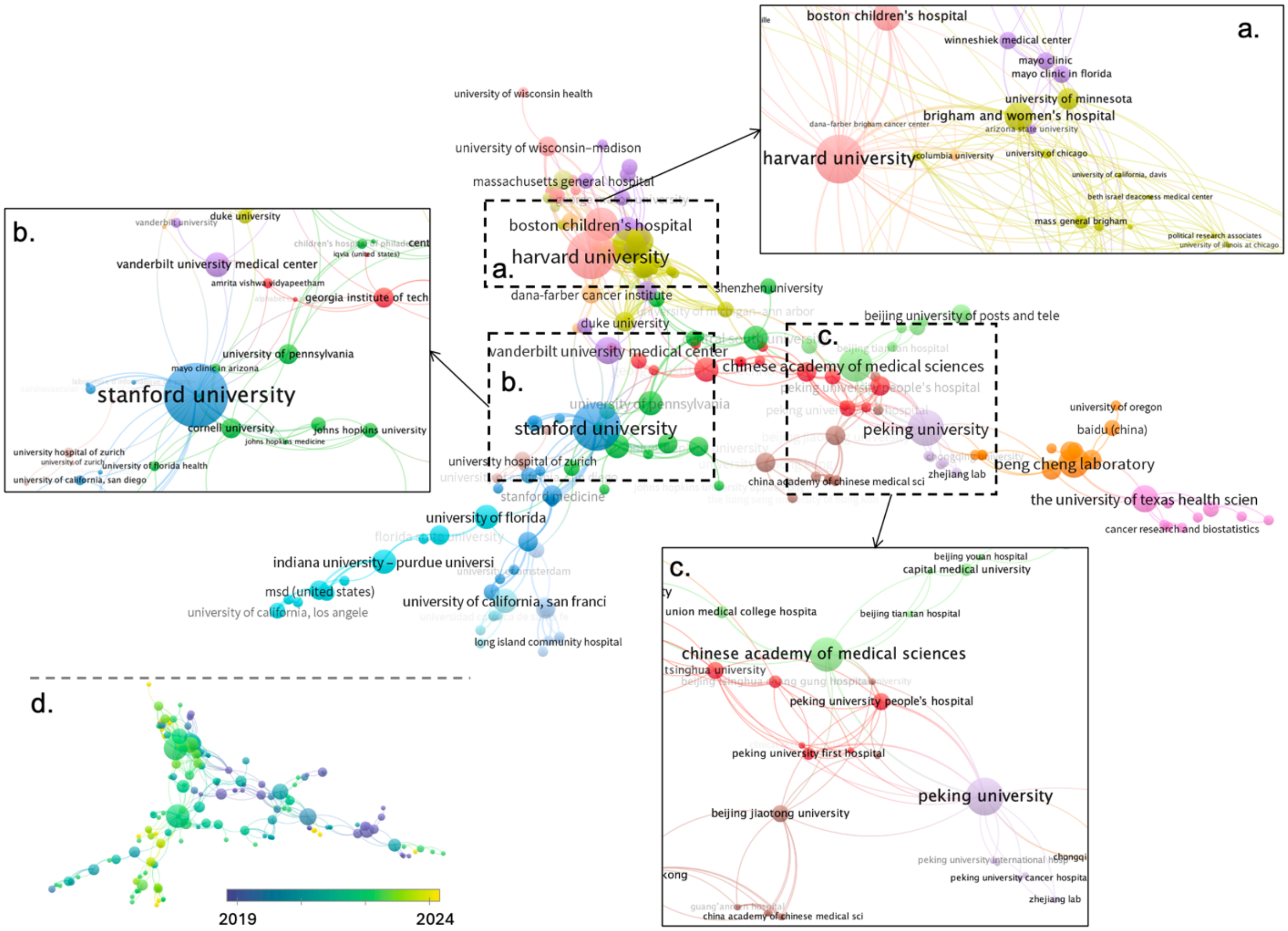

**Figure 4**. Collaboration network based on authors' affiliated institutions with (a) Cluster centered around Harvard University, (b) Cluster centered around Stanford University, (c) Cluster centered around Peking University, and (d) Collaboration overlay visualization.

## 3.2 NLP application analysis

This section focuses on the application of LLMs for each of the identified NLP tasks (excluding "other application"). First, we plotted the distribution of papers over time by each NLP task in **Figure 5**. It is noted that some studies have performed multiple NLP tasks, resulting in a cumulative total that exceeds 329. Based on **Figure 5(a)**, named entity recognition, information extraction, and diagnosis and prediction are the three most extensively investigated applications within this domain, followed by text classification and dialogue systems. By contrast, only a few studies have focused on text similarity and text summarization. One interesting observation is that three NLP tasks, including dialogue system, information extraction, and diagnosis and prediction, have seen a clear increase recently, as demonstrated by the larger circles in **Figure 5(a)**.

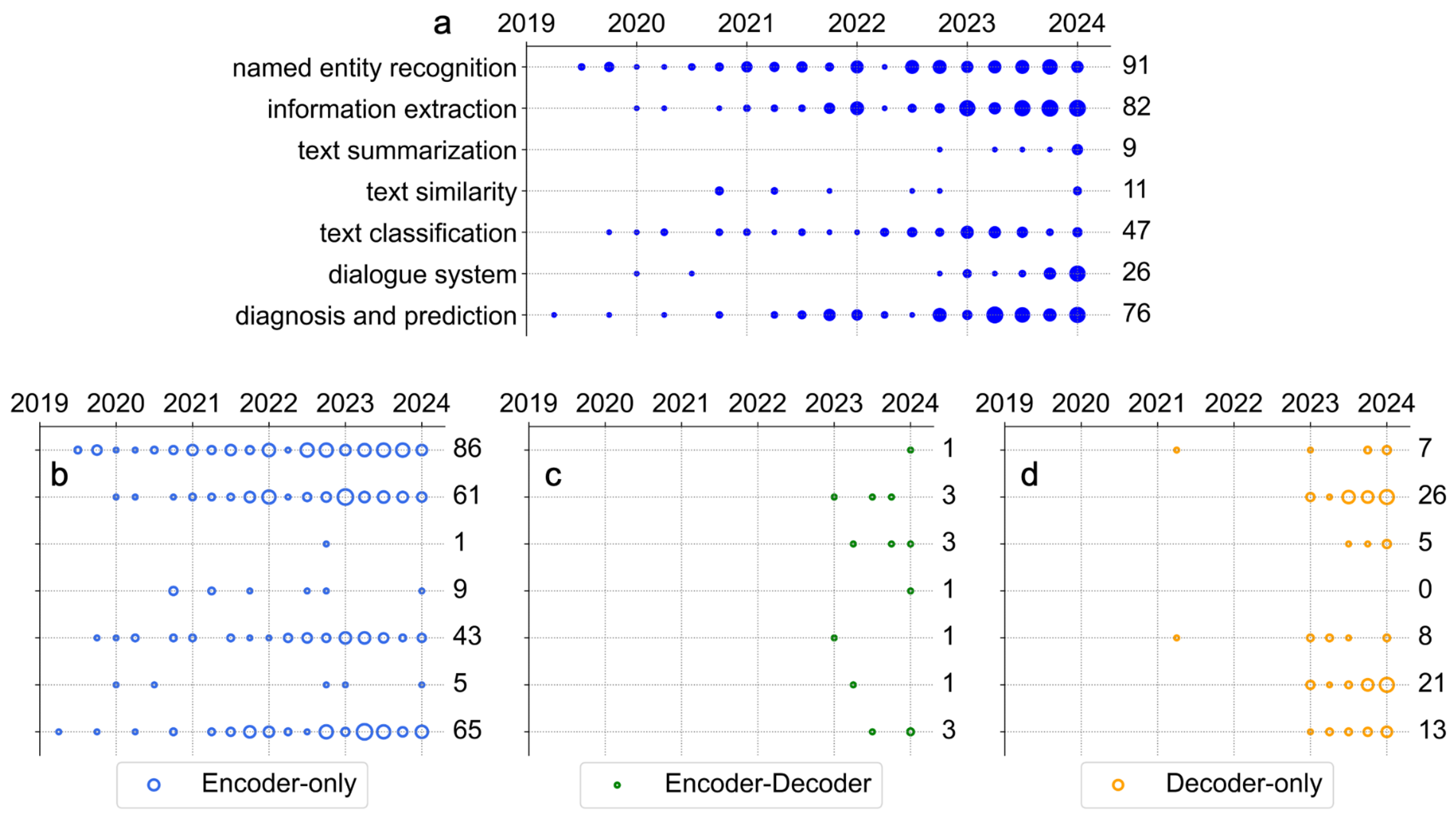

**Figure 5**. The number of studies over time is categorized by: (a) Seven identified NLP tasks, (b) NLP tasks using encoder-only models, (c) NLP tasks using encoder-decoder models, and (d) NLP tasks using decoder-only models. The x-axis shows the year, while the y-axis shows the identified NLP tasks. The size of the circle implies the number of studies.

**Figure 5(b)~(d)** presents the temporal results regarding the number of studies using encoder-only, encoder-decoder, and decoder-only models, respectively. Encoder-only models such as BERT have been extensively utilized in named entity recognition, text classification, text similarity, and diagnosis and prediction. One possible reason is that BERT generates contextualized word embeddings that capture semantic and syntactic information, making it suitable for NLP tasks when dealing with EHR data. According to **Figure 5(d)**, decoder-only models such as GPT have seen a significant increase in application across these NLP tasks following the release of ChataGPT by OpenAI in November 2022. Due to their robust language understanding and generation abilities, they have shown promise in dialogue systems, text summarization, information extraction, and diagnosis and prediction.

**Table 3**. Representative studies of using LLMs for EHR studies (performance reported in two decimal places).

| Author & year | NLP task | Research goal | Data resource | LLM | Measures & performance |
|---|---|---|---|---|---|
| Wang *et al.* (2020)[112] | Named entity recognition | Perform named entity recognition | Public dataset: CCKS2019 | BERT-BiLSTM-CRF, BERT-BiLSTMAttention-CRF | Measures: Precision, recall, F1-score<br><br>Best performance: BERT-BiLSTMAttention-CRF – F1-score of 0.89. |
| Bhate *et al.* (2023)[113] | Named entity recognition | Extract social determinants and family history | Private dataset: 1,000 narrative clinical notes from more than 150 patients with different diseases from a University hospital | GPT3.5 | Measures: Precision, recall, F1-score, accuracy<br><br>Performance: An average of 0.98 F1-score on demographics extraction, 0.62 F1-score on social determinants extraction, and 0.72 F1-score on family history extraction. |
| Zhou *et al.* (2022)[114] | Information extraction | Extract breast cancer phenotypes | Private dataset: 21 291 breast cancer patients from year 2001 to 2018 from the University of Minnesota's clinical data repository | BioBERT, CharBERT, CancerBERT | Measures: F1-score<br><br>Best performance: CancerBERT – macro F1-scores 0.88 and 0.90 for exact match and lenient match, respectively. |
| Singh *et al.* (2022)[115] | Information extraction | Link medical entity that involves mention detection and mention linking | Public dataset: i2b2-2010, MedMentions | BioBERT, UMLSBERT, BlueBERT, NeighBERT | Measures: Precision, recall, F1-score<br><br>Best performance: NeighBERT – precision 0.90, recall 0.88, and F1-score 0.89. |
| Hartman *et al.* (2023)[37] | Text summarization | Automate narrative summary of a patient's hospital stay | Private dataset: Dataset consisting of 6600 hospital admissions from 5000 unique patients at the inpatient neurology unit at NewYork-Presbyterian/Weill Cornell Medical Center | BERT, BART | Measures: ROUGE<br><br>Best performance: ROUGE scores with an R-2 of 13.76. 62% of the automated summaries meet the standard of care based on a blind-evaluation from 2 board-certified physicians. |
| Van Veen *et al.* (2023)[116] | Text summarization, text classification | Perform summarization tasks for radiology reports, patient questions, progress notes, and doctor-patient dialogue | Public dataset: MIMIC-CXR, MIMIC III, MeQSum, ProbSum, Open-i, and ACI-Bench | FLAN-T5, Med-Alpaca, Alpaca, FLAN-UL2, Llama-2, Vicuna, GPT3.5, GPT4 | Measures: BLEU, ROUGE-L, BERTScore; clinical evaluation using a five-point Likert scale regarding completeness, correctness, and conciseness;<br><br>Best performance: GPT4 achieved the best performance across these evaluations. |
| Xiong *et al.* (2020)[117] | Text similarity | Compute clinical semantic textual similarity | Public dataset: 2019 n2c2/OHNLP ClinicalSTS | BERT | Measures: Pearson<br><br>Best performance: Pearson correlation coefficient (r) of 0.85. |
| Wang *et al.* (2024)[118] | Text similarity | Standarlize obstetric diagnosis text | Private dataset: EMR data from the People's Hospital of Guangxi Zhuang Autonomous Region from April 2014 to April 2022 | BERT, ChatGLM2, Qwen | Measures: Precision, recall<br><br>Best performance: QWEN (self-consistency) – precision 0.92 and recall 0.91; MC-BERT – precision - 0.92 and recall 0.92. |
| Zhou *et al.* (2022)[119] | Text similarity | Integrate information from multiple sources to enable translations between healthcare systems | Private dataset: VA Healthcare and Mass General Brigham | BioBERT, SAPBERT, PubMed-BERT, CODER, MIKGI (their own developed model) | Measures: AUC, sensitivity, specificity<br><br>Best performance: MIKGI – AUC is 0.918 for detecting similar entity pairs and 0.809 for detecting related pairs; 0.93 sensitivity and 0.96 specificity. |
| Blanco *et al.* (2022)[120] | Text classification | Classify ICD-10 from discharge summaries | Private dataset: Swedish EHRs from Health Bank, Spanish EHRs from IXAmed-GS, Public dataset: MIMIC | BERT, PlaBERT | Measures: Precision, Recall, F1-score<br><br>Best performance: PlaBERT 0.41, 0.38, and 0.41 F1-Score on the English, Spanish and Swedish datasets, respectively |
| Chaichulee *et al.* (2022)[121] | Text classification | Multi-label classification of symptom terms from free-text bilingual adverse drug reaction reports | Private dataset: EHR from Songklanagarind Hospital | mBERT, XLM-RoBERTa, WanchanBERTa, and AllergyRoBERTa, ensemble model (aggregate) | Measures: Accuracy, hamming loss, precision, recall, F1-score<br><br>Best performance: Ensemble model – precision 0.99, recall 0.99, F1-score 0.99, and accuracy 0.98. |
| García *et al.* (2024)[122] | Dialogue system | Generate draft replies to patient inbox messages | Private dataset: compliant EHRs from a single academic medical center at Stanford Health Care | GPT3.5, GPT4 | Measures: utilization, changes in time<br><br>Performance: mean AI-generated draft response utilization rate across clinicians was 20%. There were statistically significant reductions in the 4-item physician task load score derivative. |
| Rawat et al. (2020)[123] | Dialogue system | Answer questions for EMR | Public dataset: emrQA (i2b2 challenge dataset), and MADE QA. | BERT, cBERT, cERNIE, M-cERNIE | Measures: F1-score, exact match<br><br>Best performance: M-cERNIE – 0.74 F1-score and 0.68 exact match. |
| Chung *et al.* (2024)[124] | Diagnosis and prediction | Predict perioperative risk and prognostication | Private dataset: EHR from UW Medical Center-Montlake, UW Medical Center-Northwest, Harborview Medical Center in Seattle, WA from 2021 to 2023 | GPT4 | Measures: F1-score<br><br>Performance: F1-scores of 0.50 for ASA Physical Status Classification, 0.81 for ICU admission, and 0.86 for hospital mortality. |
| Jiang et al (2023)[125] | Diagnosis and prediction | Predict readmission, in-hospital mortality, comorbidity index, length of stay, and insurance denial | Private dataset: NYU Notes containing 10 years of inpatient clinical notes for 387,144 patients. | NYUTron (their own fine-tuned BERT-like LLM) | Measures: AUC<br><br>Performance: AUC of 0.79 – 0.95 across different tasks with median AUC of 0.89. |
| Chen *et al.* (2024)[126] | Diagnosis and prediction | Identify cancer patients at risk of heart failure | Private dataset: 12,806 patients from the University of Florida Health | BERT GatorTron | Measures: Precision, recall, F1-score, AUC, accuracy.<br><br>Best performance: GatorTron-3.9B – 0.68 precision, 0.71 recall, 0.70 F1-score, and AUC 0.90. |

### 3.2.1 Named entity recognition (NER)

NER aims to recognize and classify specific types of entities present in clinical text or patient health records. The primary objective of clinical NER is to extract structured entities from unstructured text that can be easily searched, analyzed and acted upon in clinical decision-making processes. Based on our review, there are two specific use cases of NER, namely, the identification and extraction of (1) clinical terms, and (2) social determinants.

Accurate identification and classification of **clinical terms** in medical records through NER can help healthcare professionals better understand medical history, diagnoses, treatment plans, and outcomes, as well as support structured follow-up records[127]. The other application focuses on **social determinants** extraction. Social determinants in EHRs refer to the non-medical factors that can influence health outcomes[128], including a patient's socioeconomic status, education, neighborhood, employment, and access to healthcare. By identifying, extracting, and structuring social determinants, healthcare providers can gain a holistic view of a patient's health that incorporates both medical and social context. This understanding is crucial for developing more effective and personalized treatment plans.

Researchers have used various LLMs ranging from encoder-based models such as BERT to decoder-based models such as GPT to perform NER tasks. For example, Wang *et al.* proposed a BERT model called BERT-BiLSTMAttention-CRF that incorporates both lexicon features and self-attention mechanisms. The model can achieve superior performance in NER tasks based on CCKS[112]. Moreover, researchers have trained transformer-based models to extract social determinants[129,130]. Among those, BioClinical-BERT has demonstrated consistently high performance by achieving an F1-score of 0.91. However, only a few studies have attempted to use GPT-based methods to complete NER tasks[113,131]. For example, Naumann *et al.* explored the automatic extraction of social determinants using GPT4 in a one-shot prompting setting[132]. Their prompt-based GPT-4 method achieves an F1-score of 0.65 using the Social History Annotation Corpus (SHAC).

One notable trend as seen in our review is the increasing focus on NER tasks in languages other than English[133–138]. The distinct text structure and vocabulary patterns found in EHR in languages other than English suggest that standard pre-training models are less efficient at integrating entities and medical domain knowledge into representation learning. Therefore, the use of LLMs, especially those sophisticated decoder-only models such as GPT, can enhance NER tasks across different languages without the need for annotating a large EHR dataset.

### 3.2.2 Information extraction

Information extraction involves the retrieval and organization of structured information from unstructured or semi-structured EHR data. Compared to NER, it also encompasses tasks such as extracting relations, events, and attributes. Information extraction can help understand the complex interrelations and detailed events, which is crucial for effective healthcare delivery and research. It is also worth noting that some review studies consider NER as a subtask of information extraction[65]. In our study, we found that a large portion of studies (73 of 329) have only focused on NER, and therefore separated NER from information extraction.

Information extraction using LLMs can facilitate a range of advanced applications, each tailored to specific needs within healthcare. Typical applications include (1) **relation extraction** for under-standing the sequence of medical events[139,140], (2) **phenotype extraction** for detailed disease characterization[114,141], and (3) **knowledge graph construction**[142] for aggregating and visualizing medical knowledge. Taken together, these use cases demonstrate the critical role of information extraction in enhancing clinical decision-making.

Similar to NER, our review shows that BERT-based models have been extensively used for information extraction tasks. Specialized BERT models have been developed to deal with specific information extraction tasks, such as using NeighBERT[115] and CancerBERT[114] to extract relational context and cancer phenotypes, respectively. Another typical study assessed the ability of LLMs to extract adverse drug events. The study compared traditional machine learning models and LLMs such as ClinicalBERT. It revealed that ClinicalBERT outperforms traditional models in detecting drug-adverse event relations by achieving an F1-score of 0.78[143]. Other state-of-the-art LLMs, such as GPT 3.5, PaLM, LLaMa, and Flan-T5[39,144–146] have also been employed. These models often use zero- or few-shot prompts to facilitate clinical uses when annotated data is scarce or costly to obtain.

After extracting relevant entities and their corresponding relations, a knowledge graph can be created to provide a comprehensive view of the medical information content. In EHR systems, knowledge graphs can help enhance data interoperability and facilitate semantic understanding. For example, Liu *et al.*[147] established a rheumatoid arthritis knowledge graph using disease, symptoms, and treatment information extracted from a Chinese EMR. This knowledge graph offers valuable insights into various aspects related to rheumatoid arthritis, including imaging examination, affected body parts, and treatment.

### 3.2.3 Text summarization

Text summarization is important for professionals to gather information and make treatment advice, as descriptions and notes in EHRs are often lengthy and redundant. It can be applied to summarize patient's medical records or the information from a single visit, offering flexibility in addressing both comprehensive and specific healthcare needs. LLMs, due to their robust language understanding capabilities, provide significant potential for text summarization tasks. They can help condense text from various types of textual structures into a shorter concise version while retaining the key concepts and meanings from the original expressions.

The primary application of LLMs has been to **summarize patient medical information** from medical histories, progression notes, discharge notes, lab reports, image reports, and doctor-patient dialogues. Based on our review, most published articles have employed LLMs to transform discharge summaries into patient-friendly language and formats, thereby improving patient comprehension and engagement with their healthcare providers[37,148–151]. For example, Hartman *et al.*[37] utilized fine-tuned BERT and BART models to summarize EHR records and further generate discharge summaries for neurology patients who visited an academic medical center. In their study, a ROUGE (Recall-Oriented Understudy for Gisting Evaluation; a metric to evaluate automated summaries as compared to human annotators) score with an R-2 of 13.76 was reported, and 62% of the automated summaries were found to meet the standard of care under the evaluation of 2 certified physicians. LLMs have been also utilized to summarize progression notes, which incorporate a wide range of lab measurements and reports[116,152]. Van Veen *et al.*[147] presented clinical summarization using multiple LLMs, including the GPT variants, FLAN-T5, FLAN-UL2, and LlaMa, for four distinct clinical summarization tasks: radi-ology reports, patient questions, progress notes, and doctor-patient dialogues. LLM-generated summaries were equivalent (45%) or superior (36%) to medical expert summaries in terms of completeness, correctness, and conciseness with the evaluations from 10 physicians. This research illustrates that LLMs can outperform medical experts in clinical text summarization and suggests that integrating LLMs into clinical workflows could reduce documentation burdens and allow clinicians to focus more on patient care. Additionally, LLMs have been applied in summarizing question-and-answer exchanges between healthcare providers and patients, which helps to facilitate appropriate responses for health providers and promote effective interactions and communications during medical consultations[116,153].

Our review also shows that fine-tuned BERT models have been extensively utilized for text summarization, with model performance reported using correlation, F1-score, AUC, accuracy, recall, and precision metrics. BART and T5 have also been employed for summarizing data[149,152,154]. More recently, GPT4 and LlaMa-2 have been incorporated into the summarization of clinical information[148,150,151,155], with human evaluations through labeling being compared to the outputs generated by these LLMs. Compared to human labor, LLMs offer a more robust and efficient solution for summarizing EHR data due to their natural language understanding and the ability to contextualize and abstract information.

### 3.2.4 Text similarity

Text similarity in EHRs often involves identifying and quantifying the similarity between two or more pieces of clinical text data. This process aims to facilitate more efficient and accurate identification of relevant patient information, minimize data redundancy, and support clinical decision-making across multiple EHR systems[119,156]. Text similarity using LLMs is often based on semantic methods, which involve determining the similarity of two texts using their semantic embeddings.

Most studies in this area focus on concept or text matching, aiming to **enhance clinical documentation and code standardization**. For instance, Xiong *et al.*[117] introduced a semantically enriched text-matching model leveraging BERT for the ClinicalSTS dataset. Their study aimed to compute the semantic similarity among clinical text snippets. Their model could achieve a correlation coefficient of 0.85 for text snippet pair encoding. Another typical study employed text similarity to map diagnoses in EHR[118]. This study used several LLMs, including ChatGLM and Qwen for generating optimal mapping terms, and BERT for validating the performance. Their findings suggested that Qwen prompts largely outperformed the other prompts, with precision comparable to that of the BERT model.

The other promising application of text similarity is to **integrate EHR information** from multiple resources that can enable translations between healthcare systems. However, our review shows that only a few studies have been conducted in this area. Zhou *et al.*[119] proposed a multiview incomplete knowledge graph Integration algorithm using BERT to integrate information from multiple sources with partially overlapping EHR concept codes. To do so, they first generated embeddings from multiple sources, then computed the similarity between embeddings, and finally combined similar embeddings. After that, they validated the quality of combined embeddings through the tasks of mapping synonymous medication and laboratory codes across multiple health institutions.

In addition, our review shows that most current studies in this area have used BERT-based models. In comparison to traditional text vectorization techniques, such as bag-of-words (BOW), TF-IDF, and earlier word embedding techniques like FastText or Word2Vec, BERT is trained bidirectionally on large corpora, providing contextualized embeddings that take into account the specific context in which a word appears[34]. These context-sensitive capabilities make BERT particularly well-suited for text similarity tasks. Once word embeddings have been generated, researchers often choose cosine similarity to compute the similarity between vectors[156,157]. Cosine similarity calculates the cosine angle between two vectors in a multi-dimensional space, which can effectively handle high-dimensional data.

### 3.2.5 Text classification

Text classification for EHR data involves categorizing unstructured or semi-structured text data into predefined categories or classes. Given that a significant portion of EHR data is in the form of free text, such as clinical notes written by healthcare providers, text classification can help obtain meaningful insights from EHR data. However, it is important to note that text classification generally requires pre-defined categories and assigns a category to the entire text, which differentiates it from either NER or information extraction. The pre-defined categories used in text classification can vary widely depending on specific use cases, ranging from medical events[158] to disease types to patient statuses[120].

One major area focuses on classifying clinical notes into **medical domains or events**. These medical events could be medication orders, administration, drug use, and adverse reactions. Categorizing these events from EHR data allows healthcare providers to gain an understanding of patients' medication history. For example, Zhang *et al.*[159] tested several BERT-based models to classify medical aspects, including diagnosis, image inspection, laboratory, surgery, medical treatment, and anatomy from CCKS2019. Their study found that the best-performing BERT model could achieve an F1-score of 0.94. Chaichulee *et al.*[121] tested different pre-trained BERT models, including mBERT, XLM-RoBERTa, and WanchanBERTa, as well as their domain-specific AllergyRoBERTa to classify adverse drug reaction from the Songklanagarind Hospital's EHR in Thailand. According to their study, the BERT model achieved the highest performance with an F1-score of 0.99.

The second area focuses on **disease coding** classification, such as the ICD code classification. This task is about multi-label classification of health conditions, symptoms, and abnormal findings of disease based on the EHR coding system. Although this part could overlap with **Section 3.2.7**, we categorized some studies to this NLP task as they exclusively focus on text classification for ICD coding rather than patient disease diagnosis or prediction. A representative study in this area established a contextual language model with BioBERT and ClinicalXLNet techniques for ICD-10 multilabel classification[160]. They found that BioBERT had the highest F1-score of 0.70 and outperformed other models.

Our review also shows that many studies have applied BERT-based models (e.g., RoBERTa, Clinical BERT) to perform the classification tasks, with many achieving an F1-score higher than 0.8. Although encoder-based models, such as GPT-based models, are less common, those models can perform the classification tasks through fine-tuning or ICL. For example, Huang *et al.* deployed GPT 4 to identify incarceration status in the EHR data. For this binary classification, GPT4 achieved an F1-score of 0.86 through prompt design[161].

### 3.2.6 Dialogue system

The dialogue systems in healthcare, especially the question-answering scheme, can range from simple keyword searches to complex systems capable of understanding patients' or healthcare professionals' queries. These systems leverage information from medical records and generate structured text as out-put, which can enhance communication between healthcare providers and patients. In the clinical setting, dialogue systems can aid healthcare providers in **generating drafts or clarifications**[122] in response to questions regarding patients' medical history[162,163], treatment plans[164], or diagnostic findings[165]. These drafts can then be further reviewed and refined by healthcare providers before being sent to patients to ensure accurate and informed communication.

LLM-powered dialogue systems can respond to patients' questions, either with or without healthcare provider involvement. In particular, decoder-only models, due to their general-purpose abilities, can be very suitable for such application[122,162,165–169]. For example, Garcia *et al.* used GPT3.5 to generate draft replies to patient messages[122]. They found that although the mean draft response utilization rate was 20%, it could help reduce the task load for the healthcare providers. Our review also showed that many studies have developed dialogue systems to directly and automatically answer patients' questions without healthcare providers' direct involvement[40,123,162,170]. Using these LLM-aided systems, patients can directly inquire about health questions, for example, symptom inquiries[171], medication instructions and recommendations[168], in-depth explanations for medical concerns and clinical reports[172,173], or other healthcare-related queries.

Our review reveals that many LLM-based dialogue systems have been established using GPT variants, including GPT, GPT-2, GPT-3.5, and GPT-4. A key advantage of these dialogue systems built on decoder-only models is their ability to provide timely, personalized, and contextualized appropriate responses to patients' inquiries. However, it should also be noted that the dialogue systems require rigorous measures to ensure that LLMs can provide accurate, reliable, and prosperous information to patients. Our review shows that the performance of these models has been validated in the literature, as evidenced by their high scores on various evaluation metrics. For example, Vatsal *et al.* compared the GPT-generated response with the labels generated by professional medical doctors during a health plan cost-control process. They reported a mean weighted F1-score of 0.61 as compared to human estimation[38].

### 3.2.7 Diagnosis and prediction

Research to use patient EHR profiles for clinical diagnosis and risk prediction predates the popularization of large language models (see previous reviews[174–176]). The common task is to identify concurrent or future patient health-related events or risks, based on rich EHR data sources (multiple measures at multiple time points, collected for many patients). LLM-aided diagnosis and risk prediction from EHRs share many of the considerations of previous efforts, including data preprocessing (format standardization, medical code granularity), biases (missing data, loss to follow-up, and competing events), generalizability across study sites, and interpretability/ explainability of machine-based decisions. We refer to existing reviews for comprehensive discussions of these shared issues and focus on the unique opportunities enabled by LLMs in EHR-based diagnosis and prediction.

**Zero- and few-shot prompting for diagnosis/prediction.** Arguably the most straightforward way to integrate LLMs in EHR-based diagnosis and prediction is to directly query these systems with no or little additional data training. This is feasible because these models already incorporate medical knowledge in their learning[83,86]. Given the limited data-specific training, appropriate prompt design is important and requires optimization. Zhu *et al.*[177] evaluated the zero-shot performance of GPT-4 to predict various health outcomes, including in-hospital mortality, from structured EHR features in two public datasets. They reported that, with optimized prompt design, GPT-4's zero-shot prediction for in-hospital mortality outperforms common machine learning models trained on limited data. Notable prompting improvements include a) augmenting EHR test measures with their units and reference range (i.e., prior knowledge), and b) providing clinical decision-making examples as contextual anchors, which are simulated and unrelated to training and testing data but can act as pseudo reference cases to inform LLM prediction. Relatedly, Chung *et al.*[124] benchmarked the performance of GPT-4 with zero- and few-shot training across several prediction tasks using clinical notes in an in-house dataset. They found that few-shot training improves predictive performance in ICU admission and hospital mortality, whereas duration prediction tasks had universally poor performance. They also demonstrated improved performance of appropriate prompting, which involves a) providing summarized versions of the clinical notes as input, which in turn are also processed with GPT-4, and b) requesting CoT reasoning, that is, asking that the LLM provides a step-by-step explanation for its decision making. Gao *et al.*[178] examined GPT-3.5's diagnosis performance in both a public and an in-house dataset, based on patients' progress notes. They showed that improved diagnosis can be achieved with few-shot updating, and by augmenting the prompts with relevant medical concept relations as retrieved from a preconstructed medical knowledge graph. In the future, we anticipate continuous research to investigate the zero/few-shot performance of the latest LLMs in diagnosis and prediction tasks, across various patient populations, health outcomes, and EHR feature types as predictors. Importantly, we note the utility of improved prompting, as showcased in the highlighted papers, to incorporate existing medical knowledge and to enable the explainability of LMM agents' decisions.

**Longitudinal profile integration.** The longitudinal nature of EHR profiles (that various aspects of a patient's health status are captured in sequential visits) indicates that diagnosis or prediction tasks share common properties with attention-based models as adopted for sequences of tokens in language models. The architecture of LLMs can thus inherently harness the temporal dynamics of sequential EHR visits. To this end, many efforts examine the power of LLM-based models to predict patients' future health events based on their past, longitudinal profiles[126,179–182]. This is often achieved by sequentially organizing patient EHR events (e.g., diagnosis, lab tests, medication) by visiting times and providing such sequences as input to LLMs. These papers examined a variety of prediction tasks (different disease onset and prognosis, treatment response, and hospital stay duration), and found that LLM-based prediction from sequential EHR events outperformed both traditional machine learning models (e.g., random forest, gradient boosted tree) and other temporal-aware models (recurrent neural networks, long short-term memory). Of note, more works evaluated structured EHR data as the input sequence, potentially because current LLMs still have limits in the input length of tokens, rendering it infeasible to provide long sequences of unstructured clinical notes as input[183]. A recent publication[126] investigated the performance of adopting "narrative" versions of structured, longitudinal EHR profiles. Specifically, the authors evaluated the performance of a clinical LLM (GatorTron[184]) to predict the heart failure risk of cancer patients. Before feeding the patients' EHR sequence into the LLM, they first converted diagnosis and medication events into official descriptions as narrative features. They demonstrated that this improved prediction performance when compared to temporal models that did not incorporate narrative features. As LLMs are models of both sequential (temporal) attention span and word semantics, we anticipate additional future research to better leverage both aspects of longitudinal EHR profiles for improved diagnosis and prediction. Another potential future direction is to explore approaches to better account for longitudinal EHRs' time irregularities in intervals between visits, instead of inputting EHR profiles as simple sequences.

**Multimodal data integration.** Different modalities of patient EHR include unstructured (clinical notes, medical imaging) and structured (demographics, laboratory tests, clinical events, medications) domains. Many efforts to integrate across these for improved prediction with LLMs focus on combining unstructured clinical notes with structured, tabular data[185–192], with less examining the integration with medical imaging[193–195]. Across these, the common practice is to first learn each modality's specific representation and then to combine across modalities with a final, "fusion" layer for joint prediction. While state-of-the-art models can be directly adopted for individual tasks (e.g., multilayer perceptron models for encoding tabular data, transformers for clinical notes/sequential visits data, ResNet for medical images, any classification model for prediction), researchers usually make additional considerations to better adapt to EHR data types. Several efforts incorporated external biomedical knowledge (e.g., medical concept ontology and granularity, curated knowledge graphs[196]) with techniques such as attention-based representation extraction[197], graph-based representation[189], and retrieval-augmented generation[192] to clinically contextualize EHR features for improved prediction.

Another research area focuses on improved joint representation across modalities (compared to simple concatenation), especially modalities that can often be missing in sparse, irregular EHR profiles. Researchers have adopted techniques such as contrastive learning[191,193], data imputation[194], and modality-aware attention mechanisms[198] to accommodate data missingness and/or facilitate transferrable representations across modalities. While this research area has seen exciting progress, we note that recent general LLM research has further advanced towards unified multimodal LLMs[199,200], whereby non-language modality representations are aligned into text space, and, along with text data, jointly fed into an LLM "backbone" model for various tasks including prediction. This has the potential of leveraging the advanced cognitive power of state-of-the-art LLMs for multimodal applications.

# 4 Discussion

EHRs contain a wealth of patients' information and medical history. However, the challenges posed by unstructured text as well as the heterogeneity within coding systems and data formats often hinder the effective utilization of EHR data to support clinical decision-making. The development of LLMs has shown tremendous potential in language understanding and processing, with promising applications in EHR studies[130,148,201,202]. In light of these advancements, we conducted a scoping review based on 329 papers from OpenAlex focusing on the application of LLMs in EHR studies. The goal of this scoping review is to explore the current landscape in EHR research and provide useful insights for researchers and healthcare practitioners to navigate through possible challenges associated with LLM applications. Drawing insights from prior studies, we outlined several implications for researchers to consider from the perspectives of data resources, LLM techniques, performance measures, and downstream tasks in **Section 4.1**. Next, we discussed ethical concerns in **Section 4.2** and presented limitations and opportunities for future study in **Section 4.3**.

## 4.1 Practical Implications

### 4.1.1 EHR data resources

Our scoping review reveals two distinct categories of EHR data resources (**Figure 3(b)**). Each offers unique advantages. For either open-resourced databases or clinical EHRs from local hospitals and healthcare providers, it is important to note that accessing these datasets requires IRB approvals and user agreement sign-ups.

**Open-sourced databases**, such as MIMIC and CCKS, present researchers with the opportunity to study EHR data more conveniently and efficiently. Researchers can access open-sourced EHR data for research purposes without the need to collaborate with healthcare institutions. These datasets are typically well-structured and provide de-identified medical and clinical information. They have also been widely used for text classification[120], summarization[152], information extraction[203], and diagnosis and prediction analysis[204]. As such, they provide a common ground for comparing and benchmarking the performance of different LLM models. However, it is worth mentioning that these datasets, such as n2c2 and CCKS (for competition purposes), can be quite large and may not always cover specific clinical topics or patient populations of interest to researchers.

**EHRs from local hospitals and healthcare providers** present another valuable resource for researchers to consider. These datasets can be obtained from university-affiliated hospitals or clinics that provide healthcare services. Studies using these local datasets often employ samples ranging from as few as 10 to millions of EHRs[181,205,206], reflecting the diverse scale of research projects. To ensure the accuracy and relevance of these EHRs, they may require annotation by physicians or clinical professionals with expertise in relevant domains. Compared to open-resourced datasets, these EHRs allow researchers to conduct more focused investigations tailored to specific clinical topics, such as certain diseases or therapeutic prediction[124,207], and cohort studies[187]. However, obtaining access to EHRs from local healthcare institutions can be challenging, as it often requires complex ethical approval processes and data-sharing agreements.

Our findings show that fewer papers have used EHRs from open-source datasets compared to those from local hospitals. In this regard, we recommend making more open-source EHR datasets, such as MIMIC and CCKS, available to researchers. This would enable researchers who are interested in EHR studies to overcome the complex procedures often involved in accessing EHR data from local hospitals and encourage more advanced studies aimed at interpreting EHR data. To encourage the development and use of open-source EHR datasets, we also suggest policymakers consider providing funding for efforts to develop and maintain open-source databases, such as through federal grants or other funding mechanisms. This would help to ensure that these resources are available to researchers and that open-source datasets are maintained to high standards of quality and reliability.

### 4.1.2 LLM applications

**Model selection** – We suggest researchers consider model selection depending on specific EHR tasks and available computing resources. Based on **Figure 5(b)**, BERT has been extensively used in named entity recognition, text classification, text similarity, and diagnosis and prediction. Many prior studies have demonstrated the efficacy of using BERT embeddings to achieve high F1-scores in different EHR tasks[125,208,209]. One advantage of BERT is its open-source availability, which enables researchers to perform EHR tasks without incurring additional costs. In addition, compared to many open-sourced decoder-only models, such as LlaMA2, BERT can be fine-tuned with comparatively small computing resources.

Decoder-only models have demonstrated their potential in tasks such as text summarization, dialogue systems, and diagnosis and prediction (**Figure 5(d)**). Decoder-only models are particularly suitable for those EHR challenges associated with unstructured text. Even with zero-shot prompting, they can achieve high performance[177,210]. This advantage is particularly useful in situations where large sets of data annotation are not available, or the EHR data shows high heterogeneity. We also suggest researchers consider decoder-only models for the development of dialogue systems, given their general-purpose capabilities (i.e., understand users' questions and provide responses based on their understanding). These could be very useful for developing chatbots or multi-agent systems to facilitate the interactions between healthcare providers and patients[168].

**In-context learning (ICL)** – ICL is another technique associated with LLM that researchers should consider. Our scoping review found that two prompt engineering techniques have been frequently applied, including zero-shot prompting and few-shot prompting. We suggest that zero-shot prompting may be a good choice for simple tasks that do not require any specialized knowledge or training, such as information extraction from radiology reports[202], while few-shot prompting may be more appropriate for tasks that require a certain level of domain-specific knowledge, such as disease prediction[211] or ICU admission and hospital mortality prediction[124]. Other types of prompting, such as CoT prompting and self-consistency prompting have been less frequently used in current practice (a prior study applied them to text standarlization[118]), but these prompting could be useful for those problems that involve breaking down a task into a series of intermediate steps and prompting the model to generate a response for each step. This technique can be particularly useful to tackle EHR challenges that require reasoning or problem-solving, such as responses to patients' inquiries[212].

**Fine-tuning** – When fine-tuning models for EHR tasks, we recommend that researchers take into account several factors. One is the model selection based on downstream EHR tasks. For those tasks with relatively straightforward language understanding, or classifying text into structured categories such as coding classification, fine-tuning BERT models could be a cost-effective choice as they contain fewer parameters and are free to use. The second consideration is the cost especially when fine-tuning those close-source LLMs. Although the cost of some popular decoder-only models such as ChatGPT has decreased significantly[35], the upfront costs can be substantial when dealing with large-scale EHR datasets containing millions of patients' records. While open-sourced models like LLaMA2, Vicuna, and Mistral are free to use, researchers should consider the computing power required for fine-tuning these models. For example, the recommendation for fine-tuning the LLaMA2-7B model requires at least one NVIDIA A100 graphic card[86], which can be expensive for individual researchers to acquire. In addition, we suggest researchers consider several parameter-efficient fine-tuning techniques, such as adapter tuning, prompt tuning, LoRa, and QLoRa as introduced in **Section 2.3.2**. For example, Van Veen *et al.* demonstrated that QLoRa can achieve better performance than ICL in their specific text summarization tasks[116].

### 4.1.3 EHR downstream applications

Based on our scoping review, we have identified several promising directions for the downstream applications of LLMs in EHRs. One direction focuses on **enhancing the documentation for healthcare providers**.

- LLMs can facilitate the **extraction of information in a structured manner** through named entity recognition, information extraction, and text classification. LLMs can help filter out irrelevant information and standardize relevant information (e.g., diseases, symptoms, medications, and treatments) into a consistent format[213], which enables accurate interpretation as well as reduces the risk of human-made errors.

- LLMs can help generate **concise summaries of clinical notes and medical histories**[116]. The process can be either extractive, where key sentences are pulled directly from the text, or abstractive, where the information is paraphrased. Both approaches aim to condense the lengthy content while preserving its essential meaning. The summaries can be particularly helpful for healthcare providers to quickly process patients' medical history.
- LLMs can help **address the heterogeneity** presented in EHR data through ICD coding classification[160] or text similarity analysis[119]. By automating code classification and comparison, LLM can help integrate information from multiple EHR sources and enable translations between healthcare systems.

The other direction aims to **improve healthcare delivery and patient outcomes**.

- **Phenotyping** – LLMs can help support the identification and characterization of patients' traits from EHRs[214]. Accurate identification of patients' phenotyping information is helpful for personalized medicine and targeted treatment strategies.
- **Knowledge graph** – By extracting entities and their relations with the support of LLMs, LLMs can help construct knowledge graphs from EHR data. This enables the integration and visualization of complex relationships between various clinical entities, further assisting healthcare providers in making better-informed decisions.
- **Dialogue systems** – LLMs can help design chatbots responding to patients' questions or specific needs, such as drafts of emails or discharge notes[122]. In particular, LLM can help tailor the responses politely and provide customized responses, which is helpful to facilitate positive communication and ensure information is delivered appropriately.
- **Precision medicine** – Identifying patterns from a patient's EHR may indicate a certain condition or risk factor. This requires the analysis of the patient's symptoms, medical history, lab results, and other relevant information. Embedded LLMs can be utilized to predict the likelihood of a specific disease or complication, providing valuable estimations that support healthcare providers in making more accurate diagnoses[185]. This predictive capability enables clinicians to give an earlier intervention in a patient's care, potentially improving treatment outcomes.

**Multimodal integration** could be another valuable direction in EHRs. One typical example employed image-text multimodal analysis by integrating chest X-ray and EHR text to necessitate both uni-modal and cross-modal reasoning[215]. This integration of multimodal data resources can help enhance clinical decision-making and patient care. There are some typical multimodal models, such as GPT-4V[216]. In the future, we anticipate EHR-related multimodal prediction to pilot and benchmark similar unified approaches.

### 4.1.4 Performance evaluation

**Precision, Recall, F1-score, and Accuracy** have been the most commonly used evaluation metrics, particularly for tasks in NER, information extraction, text classification, and diagnosis and prediction. One consideration for using these metrics is that many annotated datasets in EHRs have imbalanced classes, such as the number of patients with a particular disease may be significantly lower than the number of patients without that disease[217], or tasks in NER and information extraction often include multiple classes with imbalanced samples[218]. In such circumstances, these metrics offer a nuanced understanding of the model's performance in each class rather than an aggregated accuracy.

**AUC, AUROC, and AUPRC** are the other set of widely used measures. However, they can only be used for binary classification unless using the one vs. all technique for multi-class classification problems. Compared to the F1-score, AUC (area under curve) and AUROC (area under receiver operating characteristic curve), which are often used interchangeably, offer an aggregate performance measure across all classification thresholds. Such a threshold-independent nature provides a comprehensive understanding of the model's prediction performance irrespective of threshold settings. The other one is the AUPRC (area under the precision-recall curve). When dealing with extremely imbalanced classes, AUPRC is often considered a more robust evaluation metric compared to AUROC, albeit less intuitive. AUPRC can serve as a more discerning metric for model selection, addressing the performance in the accurate prediction of the less common, yet significant, positive class. This is particularly useful for rare disease patient identification[219], as these are generally much less common in the EHR data.

There are also other metrics that have been used in the literature. One is the Pearson correlation coefficient. Researchers can consider using **Pearson** correlation coefficients to evaluate the semantic similarity. Another metric is **ROUGE** (Recall-Oriented Understudy for Gisting Evaluation). ROUGE and ROUGE-L are often used for text summarization[148] and dialogue systems[201]. In addition, a couple of other EHR applications, such as text summarization or dialogue systems, may require subject-matter professionals' evaluation from different dimensions, such as completeness, correctness, conciseness[154], or reading score and grade score based on the Patient Education Materials Assessment[151]. Specific explanations for these mentioned metrics are presented in **Appendix A.2**.

## 4.2 Ethical concerns

While the previous sections have shown significant potential of using LLM for EHR studies, there are several ethical concerns that researchers should keep in mind, as EHR contains extremely sensitive patient information and medication history. We briefly summarized some of them, as listed below.

- **Privacy**[220] – EHRs contain sensitive personal information. Utilizing LLMs to analyze these records raises concerns about maintaining the confidentiality and privacy of patient data. Ensuring that the data is de-identified properly before analysis and that access controls are stringent is vital.
- **Bias and fairness**[221] – LLMs can inherit and amplify biases present in their training data. These biases could affect the model's performance across different demographics, potentially leading to disparities in patient care. There is a risk that these biases can influence the model's outputs, potentially leading to unequal or unfair treatment recommendations.
- **Transparency and explainability**[222] – LLM outputs can sometimes be ambiguous, making it difficult to understand how decisions are reached. Decisions made based on LLM analysis need to be explainable, especially in a clinical setting. Healthcare providers and patients must understand how and why certain recommendations are made to trust and effectively utilize these technologies.
- **Evolving ground trut**h[223] – The ground truth in medicine is constantly evolving, making it difficult to determine whether LLMs reflect the most current data, and current models do not evaluate the quality or provide a measure of uncertainty for their outputs. Continuously updating the training data and model to reflect the latest medical knowledge is essential.
- **Accountability**[224] – Determining accountability for decisions influenced by LLM recommendations remains a challenge. Clarifying roles and responsibilities, along with developing legal and ethical frameworks, is necessary to address potential adverse outcomes linked to AI usage.

- **Ethical data use**[225] – The ethical use of EHRs extends beyond privacy. It includes ensuring that data is used in ways that respect the rights and dignity of patients. This means obtaining proper consent for the use of their data in the application of LLMs, and transparently communicating how their data will be used.

## 4.3 Opportunities for future research

There are several limitations to highlight. One limitation is associated with the paper collection process. We chose to apply the search terms to the title and abstract, which could exclude relevant papers that did not include specific keywords in these sections. Although applying search terms to the full text could potentially capture these omissions, it might introduce significant noise to the data filtering process, thereby reducing the overall accuracy of our search. Another potential limitation stems from the annotation process to classify the categories from each paper. Despite employing a pair-coding process to ensure the accuracy of our annotations, we encountered difficulties in categorizing certain NLP tasks in some studies. This may have introduced some subjectivity into our bibliometric analysis. One future work could involve more annotators with expertise to guarantee the accuracy of NLP task categorization for each paper.

In addition, it is important to acknowledge that the field of LLMs is rapidly evolving, with new research and developments emerging at an unprecedented pace. This is also the reason why we included preprinted studies for our scoping review, although some of them may not have undergone the formal peer-review process yet. While this paper provides a comprehensive discussion of the state-of-the-art LLM techniques applied in EHRs, it is inevitable that some of these methods will be surpassed by more recent advancements. As such, it is important to stay up to date with the latest developments in LLM research. Given the anticipated growth in the number of publications in the first quarter of 2024, we anticipate a publication and citation burst in the coming time. In light of this, another future work could involve monitoring this trend to ensure that our review remains at the forefront of the field.

# 5 Conclusions

The evolution of LLMs has resulted in significant advancements with diverse applications for EHR studies. In this scoping review, we have reviewed 329 relevant studies collected from OpenAlex. Our bibliometric analysis first reveals a rapid increase in research articles, particularly with the increased usage of decoder-only models, such as GPT3.5 and GPT4. Next, the institutional collaboration network shows frequent collaborations between universities and medical centers or hospitals. Further, we categorized the collected papers into seven major NLP tasks and specifically discussed the potential applications of LLMs. Our findings indicate that encoder-only models such as BERT have been extensively used in named entity recognition, information extraction, text similarity, and diagnosis and prediction, while decoder-only models such as the GPT series have illustrated promise in information extraction, dialogue system, text summarization, and diagnosis and prediction. Last, our review provides several insights for researchers to consider from the perspective of EHR data resources, LLM applications, EHR downstream applications, and performance measures. Overall, LLMs have the potential to revolutionize healthcare delivery and patient outcomes by enabling more efficient data management, improving diagnostic suggestions, enhancing patient engagement, and paving the way for personalized medicine. Several potential issues, including data privacy and AI bias, should also be addressed to ensure the responsible and ethical development and use of these technologies.

# Appendix

## A.1 The Evolution of LLMs

The table enumerates a selection of prevalent LLMs, accompanied by pertinent details including their structural design, model size, release date, and their classification as either open-source or proprietary.

**Table A.1.** The evolution of LLMs.

| Model | Architecture | Model scale | Release time | Open-source |
|---|---|---|---|---|
| GPT | decoder-only | 117M | 2018.06 | Yes |
| BERT | encoder-only | 340M | 2018.10 | Yes |
| GPT-2 | decoder-only | 1.5B | 2019.02 | Yes |
| RoBERTa | encoder-only | 355M | 2019.06 | Yes |
| ALBERT | encoder-only | 235M | 2019.09 | Yes |
| BART | encoder-decoder | 440M | 2019.10 | Yes |
| T5 | encoder-decoder | 11B | 2019.10 | Yes |
| GPT-3 | decoder-only | 175B | 2020.05 | No |
| DeBERTa | encoder-only | 1.5B | 2020.06 | Yes |
| Gopher | decoder-only | 280B | 2021.11 | Yes |
| LaMDA | decoder-only | 137B | 2022.01 | No |
| PaLM | decoder-only | 540B | 2022.04 | No |
| OPT | decoder-only | 175B | 2022.05 | Yes |
| BLOOM | decoder-only | 176B | 2022.11 | Yes |
| Galactica | decoder-only | 125B | 2022.11 | Yes |
| LLaMA | decoder-only | 65B | 2023.02 | Yes |
| GPT-4 | decoder-only | - | 2023.03 | No |
| LLaMA-2 | decoder-only | 70B | 2023.06 | Yes |
| Qwen | decoder-only | 7B | 2023.08 | Yes |
| Mistral | decoder-only | 7B | 2023.09 | Yes |
| Phi-1.5 | decoder-only | 1.3B | 2023.09 | Yes |
| Grok-1 | decoder-only | 314B | 2023.10 | Yes |
| Mixtral | decoder-only | 8x7B | 2023.11 | Yes |
| Yi | decoder-only | 34B | 2023.11 | Yes |
| Qwen | decoder-only | 72B | 2023.11 | Yes |
| Qwen-1.5 | decoder-only | 72B | 2023.11 | Yes |
| Phi-2 | decoder-only | 2.7B | 2023.12 | Yes |
| Mixtral-large | decoder-only | - | 2024.02 | No |
| Claude-3 | decoder-only | - | 2024.03 | No |
| DBRX | decoder-only | 16x8B | 2024.03 | Yes |

## A.2 Metrics for performance measures

**Precision, Recall, F-score, Accuracy** – It is worth noting these metrics, while commonly referred to as computer science terms, have corresponding traditional names in statistics. The explanation of each metric is listed below[226]. Precision (specificity) is the fraction of true positives over the number that a model classifies as positive. It is computed as:

$$Precision = \frac{TP}{TP + FP} \quad \text{Eq 1}$$

Recall (sensitivity) is the fraction of true positives over all the cases that actually are positive. It is computed as:

$$Precision = \frac{TP}{TP + FN} \quad \text{Eq 2}$$

The F1-score is a combination (the harmonic mean) of Recall and Precision. It provides a better measure of the incorrectly classified cases than the accuracy when the classes are imbalanced. Higher F1-scores imply higher performance. F1-score is computed as:

$$F1 - score = \frac{2 \times Precision \times Recall}{Precision + Recall} \quad \text{Eq 3}$$

Last, Accuracy measures the fraction of predictions that the model correctly identifies. It is computed as:

$$Accuracy = \frac{TP + TN}{TP + TN + FP + FN} \quad \text{Eq 4}$$

where the total population equals the sum of true positives (TP), true negatives (TN), false positives (FN), and false negatives (FN). Accuracy is generally computed for both the training and testing datasets for each model.

**AUC, AUROC, AUPRC** – These metrics are used for binary classification. AUC stands for "Area under the Curve." For binary classification tasks, there are two widely used curves that can measure the model's performance. One is the ROC curve (receiver operating characteristic curve), which plots the true positive rate (TPR) (y-axis) against the false positive rate (FPR) (x-axis) at each threshold setting. AUROC quantifies the entire two-dimensional area beneath the complete ROC curve, ranging from the point (0,0) to the point (1,1). The other one is the PRC curve (precision-recall curve), which measures the precision (y-axis) against recall (x-axis) for different probability thresholds. AUPRC is plotted based on precision and recall, which are more sensitive to changes in the positive class[227] (i.e., the positive class often represents the focus of interest and is less common in the dataset). As such, AUPRC places more emphasis on rare events.

**Pearson Coefficient, Spearman's Coefficient**

Pearson's correlation coefficient measures the strength of a linear relationship between two normally distributed variables. In contrast, Spearman's correlation coefficient measures the strength of a monotonic relationship between two variables that are not normally distributed, including ordinal, continuous, or contain outliers. These statistical measures help to determine the association between different variables[228].

$$Pearson\ r = \frac{\Sigma(x_i - \underline{x})(y_i - \underline{y})}{\sqrt{\Sigma(x_i - \underline{x})^2 \Sigma(y_i - \underline{y})^2}} \quad \text{Eq 5}$$

where $x_i$ and $y_i$ are the point estimations, $\underline{x}$ and $\underline{y}$ are the mean estimations.

$$Spearman\ \rho = 1 - \frac{6\Sigma d^2}{n(n^2 - 1)} \quad \text{Eq 6}$$

where d is the difference between the value of rank $x$ and $y$, and n is the number of data pairs.

**ROGUE, ROUGE-L**
ROUGE[229] compares an automatically produced summary or translation against a reference or a set of references (human-produced summaries or translations) and calculates the precision, recall, and F1-scores based on the overlapping units.

ROUGE-L[229] measures the longest common subsequence (LCS) between the candidate and reference summaries. It focuses on sentence-level structure similarity and is more suitable for evaluating longer text. Given two sequences X and Y, the LCS of X and Y is a common subsequence with maximum length.